\documentclass[letter, scriptaddress, twocolumn,  prd, showpacs,showkeys]{revtex4-1}

	\usepackage{amsmath}
           \usepackage{mathrsfs}
	\usepackage{amssymb}
	\usepackage{graphicx} 
	\usepackage{makeidx}
	\usepackage{amsfonts}
	\usepackage[ansinew]{inputenc}
	\usepackage[usenames, dvipsnames]{pstricks}
	\usepackage{epsfig}
	\usepackage{pst-grad} 
	\usepackage{pst-plot} 
	\usepackage[colorlinks, hyperindex]{hyperref}
	\hypersetup
	{
		colorlinks, %
		citecolor=blue, %
		linkcolor=Green, %
		urlcolor=blue, %
	}

\makeindex

\begin{document}

\title{{\Large\bf Restoring unitarity in anisotropic quantum cosmological models}}

\author{Sridip Pal} 
\email{sridippaliiser@gmail.com}
\altaffiliation{Present address:\\Department of Physics\\
University of California, San Diego\\
9500 Gilman Drive, La Jolla, CA 92093, USA}
\author{Narayan Banerjee}
\email{narayan@iiserkol.ac.in}
\affiliation{Department of Physical Sciences, \\
Indian Institute of Science Education and Research Kolkata, \\
Mohanpur, West Bengal 741246, India.}

\begin{abstract}
The present work shows that a properly chosen ordering of operators can restore unitarity in anisotropic quantum cosmological models. Bianchi V and Bianchi IX models with a perfect fluid are worked out. A transformation of coordinates takes the Hamiltonian to that of an inverse square potential which has equal deficiency indices; thus a self-adjoint extension is possible. Although not clearly detected clearly before, we show here that isotropic models are also apt to violate the conservation of probability for careless operator ordering.
\end{abstract}
\pacs{04.20.Cv., 04.20.Me}
\maketitle

\section{Introduction}

In the absence of a generally accepted quantum theory of gravity, quantum cosmology indeed provides an arena where signatures of quantum effects in gravitational systems are looked at. The quantum behaviour of the Universe, particularly relevant for very early stages of the evolution, is governed by Wheeler-DeWitt  equation\cite{dewitt, wheeler, misner}. However, there is a host of conceptual problems in quantum cosmology. For instance, the identification of a reasonable time parameter has indeed been a problem as in a relativistic theory time is a coordinate rather than a unequivocally respected scalar parameter \cite{kuchar1, isham, rovelli, anderson}. There are problems regarding the boundary conditions, problems regarding the interpretation of the wave function and several others. There are some comprehensive reviews which summarize the development of the subject and also the problems that it has\cite{wilt, halli, nelson1}. The problem of the identification of a well behaved time parameter might be taken care of by introducing a fluid, so that the monotonic evolution of the fluid density, if there is any, can play the role of time. This strategy had been utilized by Lapchinskii and Rubakov\cite{rubakov}. This method invokes what is known as Schutz's formalism where the fluid variables are given dynamical degrees of freedom via some thermodynamic potentials\cite{schutz1, schutz2}. The method finds a renewed and very successful application by Alvarenga and Lemos\cite{alvarenga1} and is quite frequently utilized now. For the quantization of an isotropic cosmological model, relevant examples can be found in the works of Batista et al\cite{batista}, Alvarenga et al\cite{alvarenga2} and Vakili\cite{vakili1, vakili2}. In order to quantize anisotropic models, this method has been employed by Alvarenga et al\cite{alvarenga3}, Majumder and Banerjee\cite{barun}, Pal and Banerjee\cite{sridip}.\\

A major problem in quantum cosmology is the fact that quantization of anisotropic cosmological models are notorious for the non-unitary evolution of the system leading to a non-conservation of probability. The Bianchi I model discussed by Alvarenga et al\cite{alvarenga3} and the Bianchi V and IX models investigated by Majumder and Banerjee\cite{barun} both suffer from this pathology. It should be mentioned that the use of the fluid evolution as the time parameter has the great advantage of acting as a probe for this non-unitary behaviour of the models. In the absence of matter, there is hardly any physical identification of time and thus the time dependence of probability might escape  without being properly detected\cite{lidsey, nelson2}. There is no generally accepted reason for this alleged non-unitarity, but fingers point towards the hyperbolic nature of the Hamiltonian leading to a breakdown of the positive-definiteness of the kinetic energy\cite{alvarenga3}.\\

Very recently it has been shown that, as opposed to the folklore, this non-unitarity is not in fact a generic problem of anisotropic cosmologies\cite{sridip}. The problem actually lies with a bad choice of operator ordering rather than anything else. Majumder and Banerjee\cite{barun} showed that with a proper operator ordering in the case of a Bianchi V model, the non-unitarity can be alleviated to an extent such that for large time the norm of the wavefunction becomes a constant. But one can argue that this is not quite significant as a model is either unitary or not, since unitarity is not essentially characterized by its magnitude but rather a qualitative property. But a later work\cite{sridip}, which deals with a Bianchi I model, shows that a particular operator ordering can be chosen to ensure that the Hamiltonian has a favourable deficiency index.  Thus a self-adjoint extension for the Hamiltonian is on the cards\cite{reed}. This latter work\cite{sridip} provides some explicit examples for the self-adjoint extension for the Bianchi I model and also gives the resulting solutions with a time independent norm for the wave packet. It has also been shown by the same work that a transformation of coordinates can be effected at the classical level itself so that the particular operator ordering comes naturally and does not have to be arbitrarily chosen.\\

The present work deals with examples from Bianchi V and Bianchi IX models where similar calculations lead to a unitary evolution and hence a conservation of probability. Bianchi V and IX models are of constant but nonzero spatial curvature as opposed to the zero curvature space section of the Bianchi I models. So this desired but so far eluding unitary behaviour of anisotropic quantum cosmological models is not out of reach and is also not a peculiarity of Bianchi I models alone. It is true that the solutions are obtained only for some cases of the equations of state of the fluids. But this is because of the complication in integrating the Wheeler DeWitt equation in general. However, one single non-trivial example is good enough to become hopeful, and in fact we have a host of examples now.\\

Isotropic models are believed to be well behaved and appear to present a unitary evolution. We also show that this also crucially depends on the operator ordering! It is shown that the some choice indeed yields a unitary evolution whereas there are examples in the literature\cite{alvarenga2,alvarenga3} where a different operator ordering is quite apt to present a time dependent norm. It deserves mention that the issue of operator ordering in isotropic models was raised in the context of regularity of wavefunctions much earlier by Kontoleon and Wiltshire\cite{wilt1}. \\

The paper is organized as the following. In section 2, a very brief description of the Schutz's formalism is given and the scheme of quantization for a Bianchi V model with a perfect fluid is discussed in detail. In the third section, a Bianchi IX model is discussed. The fourth section deals with a spatially flat isotropic cosmological model and the fifth and final section includes a discussion of the results obtained.

\section{Bianchi-V models}

The relevant action in gravity is given by
\begin{equation}\label{Action}
\mathcal{A}=\int_{M}d^{4}x \sqrt{-g}R +2 \int_{\partial M} \sqrt{h}h_{ab}K^{ab}+\int_{M} d^{4}x \sqrt{-g}P,
\end{equation}
where $R$ is the Ricci scalar, $K^{ab}$ is the extrinsic curvature, and $h^{ab}$ is the induced metric over the boundary $\partial M$  of the 4 dimensional space-time manifold $M$ and $P$ is the pressure of the fluid. The units are so chosen that $16\pi G = 1.$ The third integral represents the matter part taken in the form of a perfect fluid while the first two integrals take care of the gravity sector. \\

Bianchi-V cosmological models are given by the metric
\begin{equation}
\label{BianchiVmetric}
ds^{2}=n^{2}dt^{2}-a^{2}(t)dx^{2}-e^{2mx}\left[b^{2}(t)dy^{2}+c^{2}(t)dz^{2}\right]
\end{equation}
where $n(t)$ is the lapse function and $a,b,c$ are functions of the cosmic time $t$, m is a constant related to the curvature of spatial slice. This metric indicates an anisotropic but homogeneous spatial section with a constant negative curvature, determined by the constant $m$. On isotropization, this yields an open ($k=-1$) Friedmann metric. In fact $m$ can be taken to be unity without any serious loss of physical content. \\

Using the metric we rewrite the gravity sector of \eqref{Action} in the following form
\begin{equation}\label{GRaction}
\mathcal{A}_g =\int dt \left[-\frac{2abc}{n}\left(\frac{\dot{a}}{a}\frac{\dot{c}}{c}+\frac{\dot{b}}{b}\frac{\dot{a}}{a}+\frac{\dot{b}}{b}\frac{\dot{c}}{c}+\frac{3n^{2}m^{2}}{a^{2}}\right)\right].
\end{equation}
We introduce a new set of variables as
\begin{eqnarray}
\label{coordinate}
a&=&e^{\beta_{0}},\\
b&=&e^{\beta_{0}+\sqrt{3}\left(\beta_{+}-\beta_{-}\right)},\\
c&=&e^{\beta_{0}-\sqrt{3}\left(\beta_{+}-\beta_{-}\right)}.
\end{eqnarray}
This kind of the choice of variables is not new and quite extensively used in the literature \cite{alvarenga3, barun}. This choice assumes another constraint $bc=a^{2}$ which does not result in any loss of major physical properties, the model is still the anisotropic Bianchi-V, with a negative spatial curvature. This choice results in some simplification, the  Hamiltonian turns out to be free from $p_{-}$, the momentum conjugate to $\beta_{-}$ and the equations become easier to handle. With these variables, the Lagrangian density of the gravity sector becomes
\begin{equation}
\label{c4.6}
\mathcal{L}_{g}=-\frac{6e^{3\beta_{0}}}{n}\left[\dot{\beta}_{0}^{2}-\left(\dot{\beta}_{+}-\dot{\beta}_{-}\right)^{2}+e^{-2\beta_{0}}n^{2}m^{2}\right].
\end{equation}
With $\beta_{0}, \beta_{+}, \beta_{-}$ being used as the coordinates, the corresponding Hamiltonian is written as
\begin{equation}
H_{g}=-\frac{ne^{-3\beta_{0}}}{24}\left(p_{0}^{2}-p_{+}^{2}-144m^{2}e^{4\beta_{0}}\right).
\end{equation}
We now employ Schutz's formalism \cite{schutz1, schutz2} and identify a time parameter out of matter sector which is chosen to be an ideal fluid given by an equation of state $P = \alpha \rho$ where $\alpha$ is a constant (with $\alpha \leq 1$) and $\rho$ is the density of the fluid. \\

Bianchi-V differs from Bianchi-I in the proper volume measure, which is $\sqrt{-g} =\sqrt{a^{2}b^{2}c^{2}\ e^{4mx}}=e^{3\beta_{0}+2mx}$ for the former. Using standard thermodynamical considerations, the fluid part of the action \eqref{Action} can now be cast into following form
\begin{equation}
\label{3.91}
\begin{split}
\mathcal{A}_{f}&=\int dt \mathcal{L}_{f}\\&= V\int dt \left[n^{-\frac{1}{\alpha}}e^{3\beta_{0}}\frac{\alpha}{\left(1+\alpha\right)^{1+\frac{1}{\alpha}}}\left(\dot{\epsilon}+\theta\dot{S}\right)^{1+\frac{1}{\alpha}}e^{-\frac{S}{\alpha}}\right],
\end{split}
\end{equation}
where the factor of $V=\int dx\ e^{2mx}\int dy dz$ comes out due to integration over space. Here $S$, $h$ and $\epsilon$ are thermodynamic potentials. $S$ actually represents the specific entropy and $h$ is the specific enthalpy. The potential $\epsilon$ does not have any physical significance. For the details of the calculations, the meaning of the quantities used and equations connecting them, we refer to \cite{barun}.

We define the canonical momenta to be $p_{\epsilon}=\frac{\partial\mathcal{L}_{f}}{\partial\dot{\epsilon}}$ and
$p_{S}=\frac{\partial\mathcal{L}_{f}}{\partial\dot{S}}$ and Hamiltonian comes out to be
\begin{equation}
\label{hamfl}
H_{f}=ne^{-3\alpha\beta_{0}}p_{\epsilon}^{\alpha +1}e^{S}.
\end{equation}
We now effect the following canonical transformation,
\begin{eqnarray}\label{9}
T&=&-p_{S}\exp(-S)p_{\epsilon}^{-\alpha -1},\\
p_{T}&=&p_{\epsilon}^{\alpha+1}\exp(S),\\
\epsilon^{\prime}&=&\epsilon+\left(\alpha+1\right)\frac{p_{S}}{p_{\epsilon}},\\
p_{\epsilon}^{\prime}&=&p_{\epsilon},
\end{eqnarray}
and write the Hamiltonian for the fluid sector as
\begin{equation}
\label{3.93}
H_{f}= ne^{-3\beta_{0}}e^{3\left(1-\alpha\right)\beta_{0}}p_{T}.
\end{equation}
It deserves mention that the transformed variables retain the canonical structure as verified by the relevant Poisson brackets\cite{sridip}. The net Hamiltonian,  for the gravity plus the matter sector, now becomes
\begin{equation}\label{10}
H=-\frac{ne^{-3\beta_{0}}}{24}\left(p_{0}^{2}-p_{+}^{2}-144m^{2}e^{4\beta_{0}}-24e^{3\left(1-\alpha\right)\beta_{0}}p_{T}\right).
\end{equation}
Variation of the action, with respect to $n$, yields Hamiltonian constraint
\begin{equation}\label{11}
\mathcal{H}=\frac{1}{n}H=0.
\end{equation}
One should note that $T$, as defined here through the fluid variables, indeed has a proper orientation so as to be used as the time parameter\cite{sridip}. We now promote the super Hamiltonian $\mathcal{H}$ to an operator and postulate commutation relation amongst the quantum operators as usual.\\
We write
\begin{equation}\label{12}
p_{j}\mapsto -\imath\hbar\partial_{\beta_{j}},
\end{equation}
for $j=0,+$, and 
\begin{equation}\label{13}
p_{T} \mapsto -\imath \hbar\partial_{T}.
\end{equation}
This mapping is equivalent to postulating the fundamental commutation relations:
\begin{equation}\label{13.1}
\left[\beta_{j},p_{k}\right]=\imath\hbar\delta_{jk}\mathbb{I},
\end{equation}
where $\mathbb{I}$ is of course the unit matrix. In all the subsequent discussion, the choice of natural units is employed, i.e., $\hbar = 1$. With the operator ordering as prescribed in \cite{sridip}, the Wheeler-De Witt equation, $\mathcal{H}\Psi =0,$ now takes the form, 
\begin{widetext}
\begin{equation}\label{WDeq}
\left[e^{\frac{3}{2}\left(\alpha -1\right)\beta_{0}}\frac{\partial}{\partial\beta_{0}}e^{\frac{3}{2}\left(\alpha -1\right)\beta_{0}}\frac{\partial}{\partial\beta_{0}} -e^{3\left(\alpha -1\right)\beta_{0}}\frac{\partial^{2}}{\partial\beta_{+}^{2}}+144m^{2}e^{\left(3\alpha +1\right)\beta_{0}}\right]\Psi =24\imath\frac{\partial}{\partial T}\Psi.
\end{equation} 

With the standard separation of variables as, 
\begin{equation}\label{49}
\Psi(\beta_{0},\beta_{+},T)=\phi(\beta_{0})\psi(\beta_{+})e^{-\imath E T},
\end{equation}
the equation for $\phi$ becomes
\begin{equation}
\label{Operatororder}
\left[e^{\frac{3}{2}\left(\alpha -1\right)\beta_{0}}\frac{\partial}{\partial\beta_{0}}e^{\frac{3}{2}\left(\alpha -1\right)\beta_{0}}\frac{\partial}{\partial\beta_{0}} + e^{3\left(\alpha -1\right)\beta_{0}}k_{+}^{2} +144m^{2}e^{\left(3\alpha +1\right)\beta_{0}}\right]\phi =24 E\phi.
\end{equation}
\end{widetext}

This is because the solution of the $p_{+}$  sector will be of the form $e^{\imath k_{+}\beta_{+}}$ and will not be affected by the factor ordering. \\

For $\alpha\neq 1$, we make the change of variables as
\begin{equation}
\label{51}
\chi = e^{-\frac{3}{2}(\alpha-1)\beta_{0}},
\end{equation}

so that equation \eqref{Operatororder} becomes
\begin{equation}
\label{Inversesquare}
\frac{9}{4}\left(1-\alpha\right)^{2}\frac{d^{2}\phi}{d\chi^{2}}+\frac{k_{+}^{2}}{\chi^{2}}\phi +144m^{2}\chi^{\frac{2\left(3\alpha +1\right)}{3\left(1-\alpha\right)}}\phi - 24E\phi =0.
\end{equation}
We define:
\begin{eqnarray}\label{53}
\sigma &=&\frac{4k_{+}^{2}}{9\left(1-\alpha \right)^{2}},\\
E^{\prime}&=&\frac{32}{3\left(1-\alpha \right)^{2}}E.\\
M^{2}&=&\frac{64m^{2}}{\left(1-\alpha\right)^{2}}.
\end{eqnarray}
Equation \eqref{Inversesquare} can now be written as
\begin{equation}
\label{Inversesquare2}
\mathcal{H}_{g} \phi =\frac{d^{2}\phi}{d\chi^{2}}+\frac{\sigma}{\chi^{2}}\phi +M^{2}\chi^{\frac{2\left(3\alpha +1\right)} {3\left(1-\alpha\right)}}\phi  = E^{\prime}\phi.
\end{equation}
One can write \eqref{Inversesquare2} in the form
\begin{equation}
-\frac{d^{2}\phi}{d\chi^{2}}-\frac{\sigma}{\chi^{2}}\phi -M^{2}\chi^{\frac{2\left(3\alpha +1\right)} {3\left(1-\alpha\right)}}\phi = - E^{\prime}\phi.
\end{equation} 
Now \eqref{Inversesquare2} can be viewed as $-\mathcal{H}_{g}=-\frac{d^{2}}{d\chi^{2}}+V(\chi)$ with $V(\chi)=-\frac{\sigma}{\chi^{2}}-M^{2}\chi^{\frac{2\left(3\alpha +1\right)} {3\left(1-\alpha\right)}}$.  As $V(\chi)$ is a  continuous and real-valued function on the half-line for $\alpha \neq 1$, one can show that the Hamiltonian $\mathcal{H}_{g}$ has equal deficiency indices and thus admits a self-adjoint extension\cite{reed}. Although we cannot solve this equation for the general case, the standard theorem allows us to draw the conclusion. For a systematic and rigorous description of the theorem and the self-adjoint extension, we refer to the standard text by Reed and Simon\cite{reed}. So it is now proved that quantized Bianchi V model with a perfect fluid indeed admits a unitary evolution. It is important to note that this result is quite general in the sense that almost all sorts of perfect fluids are in the purview of the result. This transformation of variables, however, excludes a stiff fluid given by an equation of state $\alpha =1$. We shall now show an explicit example.

\subsection{\bf {A special case: $3\alpha +1 =0$}}

This special case appears to be a bit peculiar as the pressure is negative, but it does not violate the energy condition $\rho + 3P \geq 0$ and actually corresponds to a string distribution. The motivation behind this special choice is that it leads to a considerable simplification in the integration. In fact if we put $\alpha=-\frac{1}{3}$, the term $M^{2}\chi^{\frac{2\left(3\alpha +1\right)} {3\left(1-\alpha\right)}}$ becomes a constant and we have a similar situation like that of a Bianchi-I model with same value of $\alpha$ but with shifted energy spectra\cite{sridip} as evident from following equation
\begin{equation}\label{Master}
-\frac{d^{2}\phi}{d\chi^{2}}-\frac{\sigma}{\chi^{2}}=-\left(E^{\prime}-M^{2}\right)\phi.
\end{equation}
The solution to the \eqref{Master} has already been described in the \cite{sridip} and it is given by Hankel functions:
\begin{eqnarray}
\phi_{a}(\chi)=\sqrt{\chi}\left[AH^{(2)}_{\imath \beta}(\lambda \chi)+BH_{\imath \beta}^{(1)}(\lambda \chi)\right],\\
\phi_{b}(\chi) =\sqrt{\chi}\left[AH^{(2)}_{\alpha}(\lambda \chi)+BH_{\alpha}^{(1)}(\lambda \chi)\right],
\end{eqnarray}
respectively for $\sigma > \frac{1}{4}$ and $\sigma < \frac{1}{4}$, where  $\beta=\sqrt{\sigma-\frac{1}{4}} \in\mathbb{R}$ and $\alpha=\sqrt{\frac{1}{4}-\sigma}\in \mathbb{R}$. In both cases, the spectrum is given by 
$E^{\prime}=M^{2}-\lambda^{2}$. Hence, if we look for solutions with negative energy, we need to enforce the constraint $\lambda >M=\frac{8m}{1-\alpha}$. The self-adjoint extension guarantees that $|\frac{B}{A}|$ takes a value so as to conserve probability and makes the model unitarity\cite{sridip}.\\

For Bianchi-V cosmology with a string distribution we can summarize the effect of operator ordering via figure 1, where the norms  (or the reflection coefficient of the wave functions for a scattering like state) for various operator ordering are compared. Due to the complex nature of equations (32) and (33), we cannot make any rigorous comment on the proper volume at this stage.
\begin{figure}[!ht]
 \includegraphics[scale=0.7]{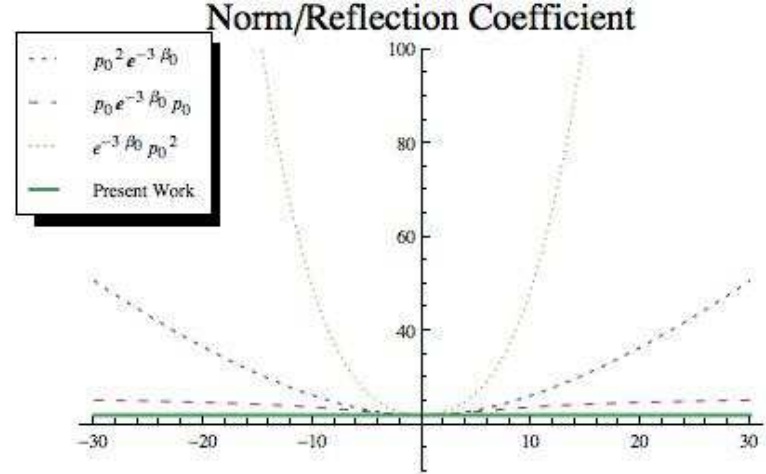}
 \caption{Behaviour of Norms (or Reflection Coefficient) against Time depending upon Particular Ordering of Operators; a suitable scaling and translation is used to enlarge the figure}
\end{figure}
Here the green one (solid thick line) is the flat one, representing a unitary model. This in fact represents reflection coefficient at origin for the solution obtained via operator ordering proposed in the present work. Being a constant (it equals one actually, the graph is scaled in order to increase the resolution in the diagram), it implies that no probability is lost at origin, hence we have a unitary model. The blue (Dashed line with medium spacing) and the red one (Dashed line with largest spacing) were obtained by Majumder and Banerjee\cite{barun}, where the red one asymptotically becomes flat and restores the probability conservation at least asymptotically. The yellow one (i.e the dashed line with tiny spacing looking like a dotted line)  (also used by Alvarenga et. al. \cite{alvarenga3} in case of Bianchi-I model) and the blue one (Dashed line with medium spacing) show clear non-unitarity as the norm is very sharply changing with time. They correspond to following operator ordering,
\begin{eqnarray*}
\textcolor{yellow}{Yellow}&\ \ &Dashed\ line\ with\ smallest\ spacing\ \equiv  e^{-3\beta_{0}}p_{0}^{2}\\
\textcolor{blue}{Blue}&\ \ &Dashed\ line\ with\ medium\ spacing\ \equiv  p_{0}e^{-3\beta_{0}}p_{0}\\
\textcolor{red}{Red}&\ \ &Dashed\ line\ with\ largest\ spacing\  \equiv  p_{0}^{2}e^{-3\beta_{0}}\\
\textcolor{green}{Green}&\ \ &Thick solid\ Line\ \ \equiv  e^{\frac{3}{2}\left(\alpha -1\right)\beta_{0}}p_{0}e^{\frac{3}{2}\left(\alpha -1\right)\beta_{0}}p_{0}
\end{eqnarray*}
So it is a gradual improvement, the red one gives the clue, a clever operator ordering makes the norm at least asymptotically time independent, whereas the present work, represented by the green one is truly unitary. But the advantage of the present work is that there is a formal proof of the fact that there exists a self-adjoint extension of the Hamiltonian. \\

\section{Bianchi-IX models}

The metric of Bianchi-IX model, which is the anisotropic generalization of the spatially closed FRW model, is given by
\begin{widetext}
\begin{equation}
ds^{2}=dt^{2}-a^{2}(t)dr^{2}-b^{2}(t)d\theta^{2}-\left[a^{2}(t)\cos^{2}(\theta)+b^{2}(t)\sin^{2}(\theta)\right]d\phi^{2}+2a^{2}(t)\cos(\theta)dr\ d\phi.
\end{equation}
\end{widetext}
Classically this metric gives a homogeneous but anisotropic space section with a constant positive curvature. On isotropization this reduces to a closed ($k=1$) Friedmann model. Bianchi IX metric has been of great utility in the investigation of the oscillatory behaviour of the universe, particularly close to the singularity. There is a rejuvenated interest in this metric in the arena of quantum cosmology, particularly supersymmetric quantum cosmology. We refer to the work of Damour and Spindel\cite{damour} and references therein in this context. \\

We define a new variable $\beta=ab$ as prescribed in \cite{barun}. The Lagrangian density for the gravity sector can be written in following form:
\begin{equation}
\mathcal{L}_{G}=\frac{2\beta^{2}\dot{a}^{2}}{a^{3}}-2\frac{\dot{\beta}^{2}}{a}-\frac{a^{5}}{2\beta^{2}}+2a.
\end{equation}
The corresponding Hamiltonian density is given by
\begin{equation}
\mathcal{H}_{G}=\frac{a^{3}p_{a}^{2}}{8\beta^{2}}-\frac{a}{8}p_{\beta}^{2}+\frac{a^{5}}{2\beta^{2}}-2a.
\end{equation}
Using Schutz's formalism along with a proper identification of time parameter as before, we come up with the Hamiltonian density for the fluid sector as
\begin{equation}
\mathcal{H}_{f}=a^{\alpha}\beta^{-2\alpha}p_{T}.
\end{equation}
The net or the super Hamiltonian for the Bianchi-IX universe is thus given by
\begin{equation}
\mathcal{H}=\mathcal{H}_{G}+\mathcal{H}_{f}=\frac{a^{3}p_{a}^{2}}{8\beta^{2}}-\frac{a}{8}p_{\beta}^{2}+\frac{a^{5}}{2\beta^{2}}-2a+a^{\alpha}\beta^{-2\alpha}p_{T}.
\end{equation}

\subsection{Ultrarelativistic Fluid $\alpha=1$}

As an example, we take up an ultrarelativistic fluid i.e $\alpha=1$ for the Bianchi IX cosmology. The motivation is the same, we can separate the Wheeler-DeWitt equation and get a complete solution. We follow the method as in \cite{barun}. For Bianchi-IX universe with ultrarelativistic fluid, the Wheeler- DeWitt equation $H\Psi = 0$ takes the form
\begin{equation}
-\frac{a^{2}}{8}\frac{\partial^{2}\Psi}{\partial a^{2}}+\frac{\beta^{2}}{8}\frac{\partial^{2}\Psi}{\partial\beta^{2}}+\left(\frac{a^{4}}{2}-2\beta^{2}\right)\Psi =\imath\frac{\partial \Psi}{\partial T}.
\end{equation}

Using separation of variable
\begin{equation}
\Psi = e^{-\imath ET}\phi(a)\psi(\beta),
\end{equation}
we have following set of equations,
\begin{eqnarray}
\label{wheeler1}
-\frac{d^{2}\psi}{d\beta^{2}}+\frac{8k}{\beta^{2}}\psi =-16\psi ,\\
\label{Wheeler2}
a^{2}\frac{d^{2}\phi}{d a^{2}}-4a^{4}\phi -8\left(k-E\right)\phi =0,
\end{eqnarray}
where $k$ is an arbitrary constant arising out of the separation process. We now define
\begin{eqnarray}
\phi &=&\frac{1}{\sqrt{a}}\phi_{0}(a),\\
\chi &=& a^{2}.
\end{eqnarray}
With this newly defined variables, we can rewrite \eqref{Wheeler2} in following fashion:
\begin{equation}
\label{wheeler2m}
-\frac{d^{2}\phi_{0}}{d\chi^{2}}-\frac{\sigma}{\chi^{2}}\phi_{0}=-\phi_{0},
\end{equation}
where $\sigma =\left[\frac{3}{16}-2\left(k-E\right)\right]$. The form of the equation \eqref{wheeler1}, \eqref{wheeler2m}, by itself ensures that the Hamiltonian has a self-adjoint extension and thereby admits a unitary evolution. Now \eqref{wheeler1}, \eqref{wheeler2m} constitute the governing equations for Bianchi-IX universe with an ultrarelativistic fluid. We observe, both of these equations can be mapped to a Schrodinger equation for a particle in an inverse square potential. Whether the potential is attractive or repulsive depends on the value of $k$ and $E$. For the attractive regime, the Hamiltonian admits a self-adjoint extension\cite{thesis,griffiths,gupta, sridip} while for repulsive regime it can be mapped to the effective equation for radial wavefunction of a free particle with energy $E_{fp}$ in 3 dimension with the identification $m=\frac{1}{2}$, $r\equiv \chi$,
\begin{equation}\label{FP}
-\frac{1}{2m}\frac{d^{2}u}{dr^{2}}+\frac{l(l+1)}{r^{2}}u=E_{fp}u.
\end{equation}
where $u(r)$ is the radial wavefunction for the free particle times the radial distance $r$ from the origin. Now in \eqref{FP}, the energy is always positive since it is not a bound state, but in \eqref{wheeler1}, \eqref{wheeler2m}, the term corresponding to $E_{fp}$ is $-16$ and $-1$ respectively, hence the repulsive regime for this case does not admit any solution. The attractive regime is enforced by following constraints
\begin{eqnarray}
k\leq 0\ \&\ 
E\leq k-\frac{3}{32}.
\end{eqnarray}\\
The inner product which makes \eqref{wheeler2m} hermitian is given by
\begin{equation}
\langle \phi_{0}|\psi_{0}\rangle \equiv \int_{0}^{\infty} d\chi\ \phi_{0}^{*}\psi_{0}=\int_{0}^{\infty}da\ 2a^{2}\phi ^{*}\psi .
\end{equation}
where $\psi=\frac{1}{\sqrt{a}}\psi_{0}$ and $\phi=\frac{1}{\sqrt{a}}\phi_{0}$.\\

Hence, for the full Hilbert space, the inner product is given by
\begin{equation}
\langle \Phi |\Psi\rangle \equiv \int_{0}^{\infty} d\beta\ \int_{0}^{\infty} da\ 2a^{2} \Phi^{*}\Psi.
\end{equation}

\section{Isotropic Homogeneous Model}

In this section we point out that without a properly chosen operator order, even an isotropic model can yield nonunitary solution. The metric for a spatially homogeneous and isotropic model with flat spatial slices is given by
\begin{equation}
ds^{2}=n^{2}(t)dt^{2}-a^{2}(t)\left[dx^{2}+dy^{2}+dz^{2}\right].
\end{equation}
The choice of operator ordering used in the present work reduces to that of Pinto-Neto et al\cite{nelson2} for the isotropic case and ensures unitarity in the model. With this operator ordering scheme, the Wheeler-DeWitt equation reads as
\begin{equation}\label{IWD0}
\left[e^{\frac{3}{2}\left(\alpha -1\right)\beta_{0}}\frac{\partial}{\partial\beta_{0}}e^{\frac{3}{2}\left(\alpha -1\right)\beta_{0}}\frac{\partial}{\partial\beta_{0}}\right]\Psi =24 \imath\frac{\partial\Psi}{\partial T},
\end{equation}
where $a=e^{\beta_{0}}$. We can easily cast this equation in terms of $a$ as,
\begin{equation}
\label{IWD}
a^{\frac{3\alpha -1}{2}}\frac{\partial}{\partial a}a^{\frac{3\alpha -1}{2}}\frac{\partial\Psi}{\partial a}=24 \imath\frac{\partial\Psi}{\partial T},
\end{equation}
which in fact conforms with that given by Pinto-Neto {\it et al}\cite{nelson2} and yields a unitary solution as shown in that work.\\

Had we not chosen this operator ordering, the solution might have been one which violates the probability conservation. For example, if we take the operator ordering prescribed in \cite{alvarenga3}, the solution will become non-unitary, which we elucidate below.\\
 
The method prescribed in \cite{alvarenga3} leads to the following Wheeler-DeWitt equation for the isotropic model (recalling that $a=e^{\beta_{0}}$ and setting $p_{\pm}=0 $ in the anisotropic Bianchi-I model, discussed in \cite{alvarenga3}):

\begin{equation}\label{IWD2}
e^{3\left(\alpha -1\right)\beta_{0}}\frac{1}{24}\frac{\partial^{2}\Psi}{\partial \beta_{0}^{2}}=\imath\frac{\partial\Psi}{\partial T},
\end{equation}
which can be shown to be not the same as \eqref{IWD}.  The equation \eqref{IWD2} admits the following definition of inner product so as to ensure the hermiticity of Hamiltonian (not the self-adjoint property though):
\begin{equation}
\langle \phi |\psi\rangle \equiv \int_{0}^{\infty}da\ a^{2-3\alpha}\phi^{*}\psi.
\end{equation}
This definition differs from that described in \cite{sridip}. Following the method described in \cite{alvarenga3}, it can further be shown that Hamiltonian defined via this operator ordering is not self-adjoint, leading to a non unitary model. \\

The bad choice of operator ordering can have an even worse effect. If we follow the operator ordering as described in \cite{alvarenga2}, for $\alpha=1$, we can not have any normalizable solution. On the other hand, the equation \eqref{IWD} or \eqref{IWD0}, following the operator ordering used in the present work, can serve the purpose right. The solution is given by $\psi = e^{\imath \left(\omega\beta_{0}+\frac{\omega^{2}}{24}T\right)}$, which can be used to construct wavepackets with time independent finite norm.\\

The point to note is that the problem of nonunitarity is more akin to a bad choice of coordinates or operator order rather than being a generic feature of anisotropic model. Here we have explicitly shown, using results from literature, that should we not employ a suitable operator ordering, the model can be plagued with non unitarity even in isotropic cases as well. \\

That the operator ordering plays an important role even in an isotropic model was shown much earlier by Konteleon and Wiltshire\cite{wilt1}. They showed that only a particular operator ordering can give rise to a consistently defined Vilenkin's tunnelling wave function\cite{vilenkin} for a spatially closed FRW model. An interesting outcome of their investigation is that the no boundary wavefunction of Hartle and Hawking\cite{hawking}, however, does not require any particular operator ordering to be defined consistently.\\

\section{Discussion}
The work conclusively shows that it is quite possible to construct unitary quantum cosmological models for anisotropic models like Bianchi V and Bianchi IX cosmologies. Along with the recent work which gives similar results for a Bianchi I model, it is now quite clear that the alleged non-conservation of probability is not really a generic problem of the anisotropy of the cosmological model, it rather results from a wrong choice of the operator ordering. The reason for choosing Bianchi V and IX is the fact that they are anisotropic generalizations of spatially open and closed isotropic models respectively. So along with the results for the Bianchi I model\cite{sridip}, one now has quite a large set of examples of anisotropic cosmologies which do preserve probability conservation on quantization. Indeed these are examples, chosen on the basis of the separability and hence the degree of simplification in the integration of the Wheeler-DeWitt equation, but in no way at the expense of anisotropy.\\

It is also shown that even isotropic models can have the same problem of a non-conservation of probability for an improper choice of operator ordering. So it is not actually fair that only anisotropic models are pulled up for the ill-behaved solutions. In some cases, explicit solutions are obtained to exhibit the unitarity of the solutions such as $\alpha= -\frac{1}{3}$ in Bianchi V and $\alpha=1$ in Bianchi IX spaces.For these cases, the Hamiltonian takes a specific form, whose self-adjoint extension has been elucidated explicitly in the reference \cite{sridip} and for a different physical context in  \cite{griffiths}. Where the solutions are not available in the closed form, one can exploit the standard theorems to check whether self-adjoint extension is possible.\\

The examples amply show that the alleged non-unitarity is not a result of the hyperbolicity of the kinetic term in the Hamiltonian. The physical basis of the operator ordering that yields a well behaved wave packet is not known. However, as no ordering starts favourite either, one can anyway choose the favourable one. \\

Certainly it would have been useful if one could arrive at a unique factor ordering. There have been attempts towards this. For example, Hall, Kumar and Reginatto\cite{hall} showed that quantizing a cosmological model in the purview of an ``exact uncertainty principle'' leads to a unique operator ordering, which amounts to $p_{0}e^{-3\beta_0}p_0$ for the present case. It has been explicitly shown\cite{barun} that this ordering leads to a nonconservation of probability in a Bianchi V model. However, it deserves mention that investigations relating to the factor ordering in quantum cosmology mostly work in the absence of an evolution of a fluid, in which case the nonunitarity can actually escape notice as there is no time marker. \\

As now one can have models which conserve probability, further physical details of quantum cosmologies for the anisotropic cases can now be investigated more thoroughly. The corresponding classical solutions should also be looked at more critically to check if there is any geodesic incompleteness or any other pathology. This is important as inverse square potential, to which the models effectively reduce, has many strange features\cite{gupta}. It has already been shown for Bianchi I models that the conservation of probability is achieved neither at the cost of the evolution of the universe nor the anisotropy itself\cite{sridip}. Now we show that it is not a peculiarity of a Bianchi I model, other anisotropic models also do have unitary evolution. The examples found in the present work, particularly for the Bianchi V models, involve Hankel functions and thus are a lot more non-trivial than their Bianchi I counterparts. So there is an ample scope for further investigation, although that is not a part of the present work.

\end{document}